\documentclass[aps,prd,groupaddress,tighten,nofootinbib,twocolumn,superscriptaddress]{revtex4}
\usepackage{mathbbold}
\usepackage{amssymb}
\usepackage{amsmath,color}
\usepackage{epsfig}
\usepackage{graphicx,epsf,epsfig}
\usepackage{bbm}

\def\a{\alpha}

\def\xa{\chi_1}
\def\xb{\chi_2}
\def\po{\phi_0}
\def\pa{\phi_1}
\def\pb{\phi_2}
\def\ka{\xi_1}
\def\kb{\xi_2}
\def\kc{\xi_3}
\newcommand{\beq}{\begin{equation}}
\newcommand{\eeq}{\end{equation}}

\begin{document}

\title{Minimal seesaw model with $S_4$ flavor symmetry }

\author{Rui-Zhi Yang}\email{ryang@mpi-hd.mpg.de}
\affiliation{Max-Planck-Institut f{\"u}r Kernphysik, Postfach
103980, 69029 Heidelberg, Germany}

\affiliation{Purple Mountain Observatory, Chinese Academy of
Sciences, Nanjing 210008, China
\\
Graduate School, Chinese Academy of Sciences, Beijing, 100012, China}

\author{He Zhang}
\email{he.zhang@mpi-hd.mpg.de}

\affiliation{Max-Planck-Institut f{\"u}r Kernphysik, Postfach
103980, 69029 Heidelberg, Germany}

\begin{abstract}
We discuss a neutrino mass model based on the $S_4$ flavor symmetry
within the minimal seesaw framework, in which only two right-handed
neutrinos are introduced and transform as ${\bf 2}$ under $S_4$.
Although the model contains less free parameters compared to the
typical seesaw models, it provides a successful description of the
observed neutrino parameters, and in particular, a nearly
tri-bimaximal mixing pattern can be naturally accommodated. In
addition, the heavy right-handed neutrino masses are found to be
non-degenerate, while only the normal hierarchical mass spectrum is
compatible with experiments for light neutrinos.
\end{abstract}

\maketitle

\section{Introduction}
\label{sec:intro}

In view of the compelling experimental evidence on neutrino
oscillations, the origin of neutrino masses and lepton flavor mixing
emerges as one of the most fundamental issues in particle physics.
Since neutrinos are massless particles in the standard model (SM) of
particle physics, a broad class of models extended the SM have been
proposed in order to accommodate massive neutrinos. The seesaw
mechanism~\cite{Minkowski:1977sc,Yanagida:1979as,GellMann:1980vs,Mohapatra:1979ia,Schechter:1980gr,Lazarides:1980nt,Mohapatra:1980yp,Foot:1988aq}
turns out to be among the most attractive extensions of the SM in
virtue of its natural explanation of tiny neutrino masses. In the
canonical type-I seesaw model, three heavy right-handed neutrinos
are introduced besides the SM particle contents, while a Majorana
mass term $M_R$ is assumed, which is not subjected to the scale of
electroweak symmetry breaking scale, i.e., $\Lambda_{\rm EW} \sim
100 ~{\rm GeV}$. The light neutrino mass scale is then strongly
suppressed with respect to $\Lambda_{\rm EW}$ due to the heavy
right-handed neutrino masses.

In general, the type-I seesaw model is pestered with too many model
parameters, and therefore, fails to predict the lepton flavor mixing
pattern as well as the light neutrino mass spectrum. For example, in
case of the simplest type-I seesaw model with three right-handed
neutrinos, there are in total fifteen free parameters in the Dirac
mass matrix together with three unknown mass eigenvalues of heavy
Majorana neutrinos, whereas the light neutrino mass matrix contains
only nine physical parameters, indicating a lack of valuable
predictions. Note that, in the most economical type-I seesaw model,
i.e., the minimal seesaw model (MSM)
\cite{Frampton:2002qc,Endoh:2002wm,Barger:2003gt,Guo:2006qa}, one
could introduce only two right-handed neutrinos, whereas the
observed neutrino mass hierarchy and lepton flavor mixing can be
well interpreted. Such a minimal extension of the SM greatly reduces
the number of free parameters, and hence is very predictive. For
instance, in the MSM, one of the light neutrinos should be massless
since $M_{R}$ is of rank 2, which indicates that $\sum_i m_i \simeq
0.05 ~{\rm eV}$ in the normal hierarchy case while $\sum_i m_i
\simeq 0.1 ~{\rm eV}$ in the inverted hierarchy case with $m_i$
being the light neutrino masses. In case that future cosmological
observations set more stringent constraints on the summation of the
light neutrino masses, the MSM would then be the most plausible
underlying model.

Recently, plenty of models extended the gauge group with flavor
symmetries are studied in order to understand the lepton flavor
mixing. In particular, the experimentally favored tri-bimaximal
mixing pattern~\cite{Harrison:2002er,Harrison:2002kp,Xing:2002sw}
can be naturally realized in many flavor symmetry models. It is
therefore interesting to investigate if the neutrino masses and
mixing can be realized in the MSM based on certain flavor
symmetries. Now that there are only two right-handed neutrinos in
the MSM, the symmetry group $G_f$ should contain at least one
two-dimensional representation, if two right-handed neutrinos are
located in the same multiplet of $G_f$. In addition, a
three-dimensional representation should be employed in order to
accommodate three generations of charged leptons in a natural way.
In this sense, the permutation group $S_4$ appears as an attractive
candidate for the MSM, since it is one of the smallest discrete
groups containing one-, two- and three-dimensional representations.
Similar models of the $S_4$ flavor symmetry within the canonical
seesaw framework have been intensively studied in the
literature~\cite{Yamanaka:1981pa,Brown:1984dk,Brown:1984mq,Ma:2005pd,Hagedorn:2006ug,Cai:2006mf,Zhang:2006fv,Koide:2007sr,Lam:2008sh,Bazzocchi:2008ej,Ishimori:2008fi,Bazzocchi:2009pv,Bazzocchi:2009da,Altarelli:2009gn,Grimus:2009pg,Dutta:2009bj,Meloni:2009cz,Ge:2010js,Hagedorn:2010th,Toorop:2010yh,Ahn:2010nw,Ishimori:2010xk,Ding:2010pc,Daikoku:2010nj,Patel:2010hr,Dong:2010zu,Ishimori:2010fs,Daikoku:2010ew,Ishimori:2010su,Park:2011zt,Ishimori:2011nv}.

In this work, we consider the MSM based on the $S_4$ flavor
symmetry. In particular, we shall show that our scheme is rather
compact whereas it is compatible with the experimental observation,
i.e., the tri-bimaximal mixing pattern could be easily accommodated.
The remaining parts of the work is organized as follows: In
Sec.~\ref{sec:sec-II}, we present the main content of our model, and
formulate the general expressions of the lepton mass matrices. One
interesting example is given in order to show how the tri-bimaximal
mixing is realized. The information on the Higgs potential is also
briefly discussed. Then, in Sec.~\ref{sec:sec-III}, we perform a
detailed numerical analysis, and illustrate the main results
obtained in the model. Finally, our conclusions are presented in
Sec.~\ref{sec:sec-IV}.

\section{The Model}
\label{sec:sec-II}

The discrete group $S_4$ is the permutation group of four distinct
objects, which contains 24 group elements and 5 irreducible real
representations. Among the five representations, two are
one-dimensional (${\bf{1_1}}$ and ${\bf{1_2}}$), one is
two-dimensional (${\bf{2}}$), and two are three dimensional
(${\bf{3_1}}$ and ${\bf{3_2}}$). The group properties, i.e., the
Kronecker products and the Clebsch Gordan coefficients, can be found
in the appendices of Ref.~\cite{Hagedorn:2006ug}.

The total symmetry of our model is then chosen to be
\begin{eqnarray}
G = SU(3)_c \otimes SU(2) \otimes U(1)_Y \otimes S_4 \, ,
\end{eqnarray}
under which the lepton content in our model is placed as
\begin{eqnarray}
L & \sim & (1,2,-1)({\bf{3_2}}) \, , \\
\ell_R & \sim & (1,1,-2)({\bf{3_2}}) \, , \\
\nu_R & \sim & (1,1,0)({\bf{2}}) \, .
\end{eqnarray}
Note that in Ref.~\cite{Park:2011zt}, the minimal seesaw model is
considered whereas the two right-handed neutrinos are assigned to
the trivial representation of $S_4$. Furthermore, the Higgs
assignments in our model are given by
\begin{eqnarray}
\phi_0 & \sim & (1,2,-1)({\bf{1_1}}) \, , \\
(\phi_1,\phi_2) & \sim & (1,2,-1)({\bf{2}}) \, , \\
(\xi_1,\xi_2,\xi_3) & \sim & (1,2,-1)({\bf{3_1}}) \, , \\
(\chi_1,\chi_2) & \sim & (1,1,0)({\bf{2}}) \, ,
\end{eqnarray}
where the $SU(2)$ doublets Higgs fields are in analogy to these in
Ref.~\cite{Hagedorn:2006ug}, whereas an additional $SU(2)$ singlet
Higgs $\chi$ is introduced. We will show later on that $\chi$ is
crucial to ensure the corrected prediction on the neutrino mixing
angles as well as the light neutrino masses. Note that we mainly
focus our attention on the lepton flavor mixing, and hence do not
include the quark sector in our discussions. A simple way to contain
the quark mixing in our model is to make a naive assumption that all
the quarks belong to the identity representation, i.e., ${\bf 1_1}$,
and then the quark flavor mixing and masses can be obtained via the
standard Yukawa couplings to $\phi_0$.

By using the group algebra of $S_4$, we can write the invariant
Yukawa couplings for leptons as
\begin{eqnarray}
{\cal L} & = & \a_0 \left(\overline{L_1} e_R+ \overline{L_2}\mu_R+
\overline{L_3}\tau_R \right)\phi_0 \nonumber
\\ & +& \a_1 \left[ \sqrt3 \left(\overline{L_2}\mu_R-\overline{L_3}\tau_R \right)\phi_1 \right. \nonumber \\
&&
\left.+(-2\overline{L_1}
e_R+\overline{L_2}\mu_R+\overline{L_3}\tau_R)\phi_2\right] \nonumber
\\ & +& \a_2 \left[\left(\overline{L_2}\tau_R + \overline{L_3}\mu_R\right)\xi_1
+ \left(\overline{L_1}\tau_R + \overline{L_3} e_R\right)\xi_2
\right. \nonumber
\\
&&+ \left. \left(\overline{L_1} \mu_R + \overline{L_2} e_R
\right)\xi_3 \right]
\nonumber \\
&+& \beta_0 \left[ \frac{2}{\sqrt{6}}
\overline{L_1}\nu_{R1}\tilde{\xi_1}+\left(-\overline{L_2}\nu_{R1}+\sqrt{3}\overline{L_2}\nu_{R2}\right)\tilde{\xi_2}
\right. \nonumber \\ &&+ \left. \left(-\overline{L_3}\nu_{R1}
-\sqrt{3} \overline{L_3}\nu_{R2}\right)\tilde{\xi_3} \right]
\nonumber
\\
&+&  \frac{\beta_1}{2}  \left[\left(\overline{\nu^c_{R1}}\nu_{R2}+
\overline{{\nu}^c_{R2}}\nu_{R1}\right)\chi_1 +
\left(\overline{{\nu}^c_{R1}}\nu_{R1} -
\overline{{\nu}^c_{R2}}\nu_{R2}\right)\chi_2\right]
\nonumber \\
&+& \frac{M}{2}\left(\overline{\nu_{R1}^c}\nu_{R1} +
\overline{\nu_{R2}^c}\nu_{R2}\right) + {\rm h.c.} \, ,
\end{eqnarray}
where $\tilde{\xi_i}$ is the conjugate of $\xi_i$ related by
$\tilde{\xi_i} \equiv {\rm i} \tau_2 \xi^*_i$, and a bare Majorana
mass $M$ is included.

\subsection{Charged lepton masses}

In our model, the $S_4$ flavor symmetry is assumed to be
spontaneously broken by the vacuum expectation values (VEVs) of
Higgs scalars, i.e., $\langle \phi_i \rangle=v_i$,
$\langle\xi_i\rangle=u_i$, and $\langle\chi_i\rangle=x_i$. One then
arrives at the mass matrix of charged leptons as
\begin{eqnarray}\label{eq:Ml}
 M_\ell = \left(
\begin{matrix}
 a_0-2a_2 & b_3 & b_2
\cr b_3 & a_0+\sqrt{3}a_1+a_2 & b_1 \cr b_2 & b_1 &
a_0-\sqrt{3}a_1+a_2
\end{matrix}
\right) \, ,
\end{eqnarray}
where we have defined $a_0 =\alpha_0 v_0$, $(a_1,a_2)=(\alpha_1 v_1,
\alpha_1 v_2)$, and $b_i=\alpha_2 u_i$ (for $i=1,2,3$). In general,
all the parameters in the mass matrix can be complex, while in case
of CP-conservation, there are totally six real parameters in
$M_\ell$. For simplicity, we will take all the parameters to be
real, but comment later on the most general case with CP-violating
effects.

According to Eq.~\eqref{eq:Ml}, the contributions from $\phi_i$
merely affect the diagonal entries, whereas $\xi_i$ appear in the
off-diagonal elements. In the limit $a_i \gg b_i$, $M_\ell$
approximates to a nearly diagonal form, and the charged-lepton
masses are solely determined by $a_i$. Note that this is indeed a
very realistic scenario if the VEVs of $\xi_i$ are much smaller than
those of $\phi_i$. Explicitly, the sum of the VEVs has to be equal
to the electroweak scale, i.e., $\sum_i \left|{\rm VEV}_i\right|^2
\simeq \left( 174 ~{\rm GeV}\right)^2$. Since $\phi_0$ should also
be responsible for the generation of the top-quark mass, one may
reasonably take $v_0 \simeq 174 ~{\rm GeV}$ with all the other VEVs
being much smaller than $v_0$. In our model, we assume that $v_1,
v_2 \sim {\rm GeV}$ and $u_i \sim {\rm MeV}$. As a result, the
eigenvalues of $M_\ell$ are approximately given by $ a_0-2a_2$,
$a_0+\sqrt{3}a_1+a_2$, and $a_0-\sqrt{3}a_1+a_2$, respectively.
Compared to the charged-lepton masses, one immediately obtains
\begin{eqnarray}\label{eq:Me}
a_0 & \simeq & \frac{1}{3} \left( m_e + m_\mu + m_\tau\right) \, ,
\\
a_1 & \simeq & \frac{1}{2\sqrt{3}} \left(  m_\mu - m_\tau \right) \, ,  \\
a_2 & \simeq & \frac{1}{6} \left(  m_\mu + m_\tau -2m_e\right) \, .
\end{eqnarray}
In addition, the diagonalization matrix for $M_\ell$ is nearly an
identity matrix, i.e., $V_\ell \simeq I$.

\subsection{Neutrino mass matrix}

Since there are only two right-handed neutrinos in the MSM, the
Dirac mass of neutrinos is a $3\times 2$ matrix, viz.
\begin{eqnarray}
M_D= \left(
\begin{matrix}
 2X_1 & 0
\cr -X_2 & \sqrt{3}X_2 \cr -X_3 & \sqrt{3}X_3
\end{matrix}
\right) \, ,
\end{eqnarray}
where $X_i=\frac{\beta_i u_i}{\sqrt{6}}$ for $i=1,2,3$. The
right-handed neutrino mass matrix in our model is given by
\begin{eqnarray}\label{eq:MR}
M_R= \left(
\begin{matrix}
 A+C & B
\cr B & A-C
\end{matrix}
\right) \, ,
\end{eqnarray}
where $A=M_1$, $B=\beta_1 x_1$, and $C=\beta_1 x_2$, respectively.
In case of $M_R \gg M_D$, Eq.~\eqref{eq:MR} leads to the masses of
right-handed neutrinos as
\begin{eqnarray}
M_{1,2} = A \pm \sqrt{B^2 + C^2} \, .
\end{eqnarray}
Note that, as aforementioned, the natural scale of $M_D$ relies on
the VEVs $u_i$ implying $X_i \sim {\cal O}({\rm MeV})$. This in turn
helps us to estimate that the right-handed neutrino masses should be
around ${\cal O}(10^2) ~{\rm TeV}$, which turn out to be beyond the
scope of forthcoming collider experiments. In case that certain
fine-tuning is involved in the seesaw formula, e.g., the structural
cancellation, one can, at least in principle, bring the masses of
right-handed neutrinos down to the electroweak scale, although the
naturalness of such low-scale right-handed neutrinos seems
questionable.\footnote{Note that, the realization of the TeV minimal
seesaw model turns out to be more natural compared to the typical
low-scale type-I seesaw model, since the light neutrino masses could
be protected by certain underlying symmetries and hence do not
suffer from large radiative corrections~\cite{Zhang:2009ac}.}

By using the standard seesaw formula, i.e., $m_{\nu}=-M_D M_R^{-1}
M_D^T$, we obtain the light neutrino mass matrix as
\begin{eqnarray} \label{eq:mnu}
m_\nu  = m_0 \times \qquad\qquad\qquad\qquad\qquad \qquad\qquad\qquad\qquad\nonumber  \\
\left(
\begin{matrix}
 2\epsilon_2 - 2 & (1 + \sqrt{3} \epsilon_1 - \epsilon_2) r_1 &  (1 - \sqrt{3} \epsilon_1 - \epsilon_2)
 r_2
\cr \sim & -(2 + \sqrt{3} \epsilon_1 + \epsilon_2) r^2_1 & (1 + 2
\epsilon_2) r_1 r_2 \cr \sim & \sim & (\sqrt{3} \epsilon_1 -
\epsilon_2 -2) r_2^2
\end{matrix}
\right) \, ,
\end{eqnarray}
where
\begin{eqnarray}
m_0 = \frac{2 X^2_1 A}{(A^2 - B^2 - C^2)} \, ,
\end{eqnarray}
and the parameters $\epsilon$ and $r$ are defined by $\epsilon_1 =
B/A$, $\epsilon_2 = C/A$, $r_1 = X_2/X_1$, and $r_2 = X_3/X_1$.

\subsection{Lepton flavor mixing}

The light neutrino mass matrix $m_\nu$ is symmetric, and thus can be
diagonalized by means of a unitary matrix $V_\nu$ as $V^\dagger_\nu
m_\nu V^*_\nu = {\rm diag} (m_1,m_2,m_3)$. The lepton flavor mixing
matrix $U$ which links the neutrino mass eigenstates with their
flavor eigenstates is then given by
\begin{eqnarray}
U=V_{\ell}^{\dagger}V_{\nu} \simeq V_{\nu} \, ,
\end{eqnarray}
where the last approximation follows since we have taken the
charged-lepton mass matrix to be nearly diagonal. In the standard
(i.e., CKM-like) parametrization one has
\begin{eqnarray}\label{eq:parametrization}
U  = R_{23}P_{\delta}R_{13}P_{\delta}^{-1}R_{12}P_{M} \ ,
\end{eqnarray}
where $R_{ij}$ correspond to the elementary rotations in the
$ij=23$, $13$, and $12$ planes (parametrized in what follows by
three mixing angles $c^{}_{ij} \equiv \cos \theta^{}_{ij}$ and
$s^{}_{ij} \equiv \sin \theta^{}_{ij}$), $P_{\delta}={\rm
diag}(1,1,{\rm e}^{{\rm i}\delta})$, and $P_{M}={\rm diag}({\rm
e}^{{\rm i}\alpha_{1}/2}, {\rm e}^{{\rm i}\alpha_{2}/2},1)$ contain
the Dirac and Majorana CP-violating phases, respectively.

In order to get the explicit expression of $U$, a fully
diagonalization of $m_\nu$ is involved, and the results are rather
tedious. However, since the $m_\nu$ is of rank 2, there exists a
eigenvector $\bar{k} = (r_1 r_2,r_2,r_1)^T$ satisfying $m_\nu
\bar{k} =0$. In case that the light neutrino mass spectrum is
inverted hierarchy (i.e., $m_2 > m_1 \gg m_3$), $\bar{k}$
corresponds to the third column of $U$. Compared to
Eq.~\eqref{eq:parametrization}, we obtain
\begin{eqnarray}
\tan\theta_{23} & = & \frac{r_2}{r_1} \, , \\
\tan\theta_{13} & = & \frac{r_1 r_2}{\sqrt{r^2_1 + r^2_2}} \, .
\end{eqnarray}
In view of the experimentally measured maximal atmospheric angle and
small reactor mixing angle, the relation $r_1 \simeq r_2 \ll 1$ has
to be fulfilled. The two non-vanishing masses are then approximately
given by
\begin{eqnarray}
m_1 & \simeq & m_0 \left[ 1-\epsilon_2 + \sqrt{(1-\epsilon_2)^2}
\right] + {\cal O}(r_1,r_2) \, , \\
m_2 & \simeq & m_0 \left[ 1-\epsilon_2 - \sqrt{(1-\epsilon_2)^2}
\right] + {\cal O}(r_1,r_2)\, .
\end{eqnarray}
No matter what value of $\epsilon_2$ one chooses, it is not possible
to let the two masses to be nearly degenerate (i.e., $m_1 \simeq
m_2$), which is indeed required for the inverted mass hierarchy
case. Therefore, by analyzing the eigenvector of $m_\nu$, we can
conclude that the inverted light neutrino mass hierarchy is not
compatible with the model.

Henceforth, we shall concentrate on the normal hierarchy case,
namely $m_1<m_2 \ll m_3$. Here, we show one interesting example, in
which the tri-bimaximal mixing pattern (i.e., $\theta_{12}\cong
35.3^\circ$, $\theta_{23} = 45^\circ$ and $\theta_{13}=0$) is
predicted. Concretely, we make the assumptions that $r_1 = r_2 = 2$
and $\epsilon_1=0$. Equation \eqref{eq:mnu} now reduces to
\begin{eqnarray} \label{eq:mnu}
m_\nu  = m_0 \left(
\begin{matrix}
2\epsilon_2 - 2 & 2 (1 - \epsilon_2) &  2 (1 - \epsilon_2) \cr \sim
& -4(2  + \epsilon_2)  & 4 (1 + 2 \epsilon_2)  \cr \sim & \sim & -4
(\epsilon_2 +2)
\end{matrix}
\right) \, .
\end{eqnarray}
One observes from Eq.~\eqref{eq:mnu} that, with the assumptions
above, a $\mu-\tau$ symmetry appears in $m_\nu$, which generally
predicts a maximal atmospheric mixing angle, i.e., $\theta_{23} =
45^\circ$, and a vanishing $\theta_{13}$. It is then easy to prove
that the diagonalization matrix of $m_\nu$ takes exactly the
tri-bimaximal mixing form, i.e.,
\begin{eqnarray}
U_{\rm TB} = \left(
\begin{matrix}
\sqrt{\frac{2}{3}} & \sqrt{\frac{1}{3}} & 0 \cr \sqrt{\frac{1}{6}} &
-\sqrt{\frac{1}{3}} &-\sqrt{\frac{1}{2}} \cr \sqrt{\frac{1}{6}} &
-\sqrt{\frac{1}{3}} & \sqrt{\frac{1}{2}}
\end{matrix}
\right) \, ,
\end{eqnarray}
while the light neutrino masses are given by
\begin{eqnarray}
m_1 & = & 0 \, , \\
m_2 & = & 6 m_0 (1 - \epsilon_2) \, , \\
m_3 & = & 12 m_0 (1 + \epsilon_2)  \, .
\end{eqnarray}
Consequently, both the tri-bimaximal mixing and the normal neutrino
mass spectrum ($m_1<m_2 \ll m_3$) are accommodated.

Furthermore, if we relax the assumptions on $r_i$ and $\epsilon_i$,
a deviation from the tri-bimaximal mixing can be achieved. Fox
example, in the case $\epsilon_1 \neq 0$, the light neutrino mass
matrix can be written as
\begin{eqnarray}
 m_\nu  &  = &  m_0 U_{\rm TB} \left(
\begin{matrix}
0 & 0 & 0 \cr 0 & 6 (\epsilon_2 -1) & -6 \sqrt{2} \epsilon_1 \cr 0 &
\sim & -12 (1 + \epsilon_2)
\end{matrix}
\right) U^T_{\rm TB} \nonumber \\
&= & U_{\rm TB} R_{23}(\theta) ~ {\rm diag} (0,m_2,m_3)
~R^T_{23}(\theta) U^T_{\rm TB} \, ,
\end{eqnarray}
with
\begin{eqnarray}
m_2 & = & 3 m_0 \left( 3 + \epsilon_2 - \sqrt{(1+3\epsilon_2)^2 + 8 \epsilon^2_1 }\right) \, , \\
m_3 & = & 3 m_0 \left( 3 + \epsilon_2 + \sqrt{(1+3\epsilon_2)^2 + 8
\epsilon^2_1 }\right)  \, ,
\end{eqnarray}
and
\begin{eqnarray}
\sin2\theta = \frac{2\sqrt{2}\epsilon_1}{\sqrt{(1 + 3 \epsilon_2)^2
+ 8 \epsilon^2_1 }} \, .
\end{eqnarray}
The neutrino mixing angles are then modified to
\begin{eqnarray} \label{eq:s12}
s_{12} & = & \frac{1}{\sqrt{3}} - \frac{2}{\sqrt{3}} \sin^2\frac{\theta}{2} \, ,   \label{eq:s23} \\
s_{23} & = & \frac{1}{\sqrt{2}} + \frac{1}{\sqrt{3}} \sin\theta -
\frac{2}{\sqrt{2}} \sin^2\frac{\theta}{2} \, , \\
s_{13} & = & \frac{1}{\sqrt{3}} \sin\theta \, .  \label{eq:s13}
\end{eqnarray}
According to the above equations, the deviations of $\theta_{ij}$
from their exact tri-bimaximal values are correlated by $\theta$,
and in the limit $\theta \to 0$ (or effectively $\epsilon_1 \to 0$),
the exact tri-bimaximal mixing will be reproduced. Note that, the
correction to $s_{12}$ is proportional to $\sin^2\frac{\theta}{2}$,
and thus is strongly suppressed for a small $\theta$. Therefore,
$\theta_{12}$ is rather stable against $\epsilon_1$
corrections~\cite{Albright:2008rp,Albright:2010ap}.

\subsection{Higgs potential}

Now that the previous discussions rely on the VEVs of the scalar
fields, we are coming to the question of the possible Higgs
potential and its minima. Apart from the $SU(2)$ singlets $\chi_i$
our Higgs setup is essentially the same as the Higgs sector
considered in Ref.~\cite{Hagedorn:2006ug}, where only $SU(2)$
doublets are introduced. We thereby only show the Higgs potential
parts involving $\chi_i$, viz.,
\begin{widetext}
\begin{eqnarray}
V_{\chi}&=&-\mu^2_1\left(\xa^2+\xb^2\right)+\mu_2\left(3\xa^2\xb-\xb^3\right) + \omega_1\left(\xa^2+\xb^2\right)^2 + \omega_2\left[\left(\xa\xb+\xb\xa\right)+\left(\xa^2-\xb^2\right)\right]^2 \nonumber\\
&+&\rho_1\left[\po^{\dagger}\po\left(\xa^2-\xb^2\right)\right]+\rho_2\left(\left|\po^{\dagger}\xa\right|^2+\left|\po^{\dagger}\xb\right|^2\right) + \rho_3\left(\po^{\dagger}\pa\xa+\po^{\dagger}\pb\xb+{\rm h.c.}\right) \nonumber \\
&+&\rho_4\left[\po^{\dagger}\pa\left(\xa\xb+\xb\xa\right)+\po^{\dagger}\pb\left(\xa^2-\xb^2\right)+{\rm h.c.}\right] \nonumber \\
&+&\varepsilon_1 \left(\pa^{\dagger}\pa+\pb^{\dagger}\pb \right)\left(\xa^2+\xb^2\right) + \varepsilon_2 \left[\left(\pa^{\dagger}\pb+\pb^{\dagger}\pa\right)\left(\xa\xb+\xb\xa\right) + \left(\pa^{\dagger}\pa-\pb^{\dagger}\pb\right)\left(\xa^2-\xb^2\right)\right] \nonumber \\
&+&\varepsilon_3\left[\left(\pa^{\dagger}\pb+\pb^{\dagger}\pa\right)\xa+\left(\pa^{\dagger}\pa-\pb^{\dagger}\pb\right)\xb\right]+ \varepsilon_4\left|\pa^{\dagger}\xa+\pb^{\dagger}\xb\right|^2+\varepsilon_5\left|\pa^{\dagger}\xb+\pb^{\dagger}\xa\right|^2 \nonumber \\
&+&\varepsilon_6\left(\left|\pa^{\dagger}\xb+\pb^{\dagger}\xa\right|^2+\left|\pa^{\dagger}\xa-\pb^{\dagger}\xb\right|^2\right) \nonumber \\
&+&k_1\left(\ka^{\dagger}\ka+\kb^{\dagger}\kb+\kc^{\dagger}\kc\right)\left(\xa^2+\xb^2\right) + k_2\left[\sqrt{3}\left(\kb^{\dagger}\kb-\kc^{\dagger}\kc\right)\left(\xa\xb+\xb\xa\right) +\left(\kb^{\dagger}\kb+\kc^{\dagger}\kc-2\ka^{\dagger}\ka\right)\left(\xa^2-\xb^2\right)\right] \nonumber \\
&+&k_3\left(4\left|\ka \xa\right|^2+\left|\sqrt{3}\kb \xa+\kb\xb\right|^2+\left|\sqrt{3}\kc\xa-\kc\xb\right|^2\right) + k_4\left(4\left|\ka\xa\right|^2+\left|\sqrt{3}\kb\xb-\kb\xa\right|^2+\left|\sqrt{3}\kc\xb+\kc\xa\right|^2\right) \nonumber \\
&+&k_5\left[\sqrt{3}\left(\kb^{\dagger}\kb-\kc^{\dagger}\kc\right)\xa+\left(-2\ka^{\dagger}\ka+\kb^{\dagger}\kb+\kc^{\dagger}\kc\right)\xb\right]
\, .
\end{eqnarray}
\end{widetext}
Compared to the Higgs potential in Ref.~\cite{Hagedorn:2006ug},
there are in total 21 more parameters. So we are confident to arrive
at the suitable minima of the Higgs potential, and the VEV structure
described in the previous analysis can be easily satisfied.
Furthermore, we did not discuss in detail the Higgs spectrum, which
may result in the flavor changing neutral currents as well as lepton
flavor violating problems. However, such problems commonly occur in
models with more than one Higgs doublets, and can be ignored if the
flavor changing Higgs are all heavier than a few TeV.

\section{Numerical illustrations}
\label{sec:sec-III}

\begin{figure}[t]\vspace{-1.cm}
\includegraphics[width=90mm,bb=80 -20 430 330]{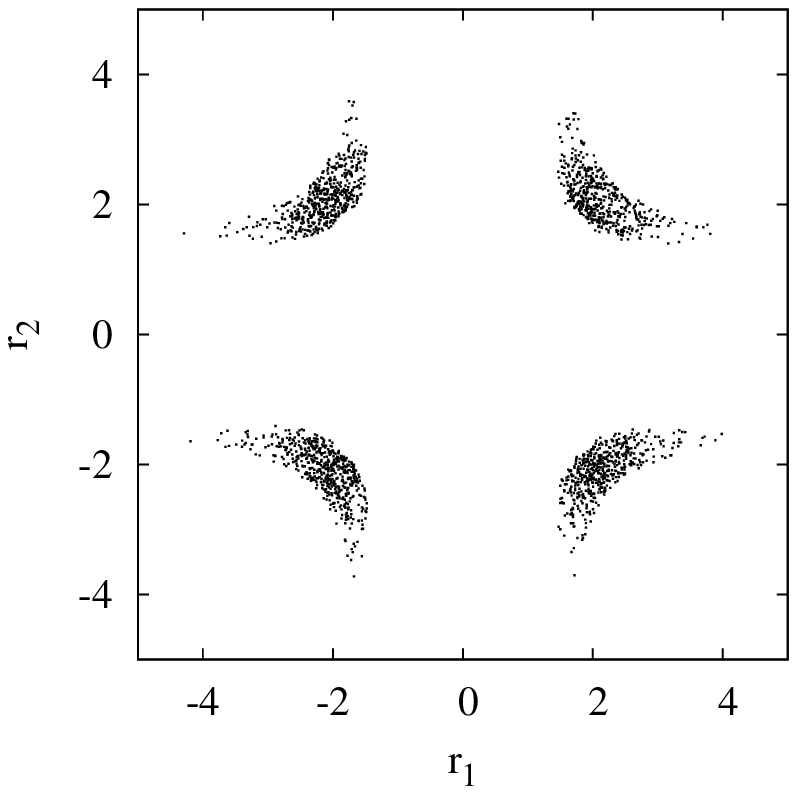}
\includegraphics[width=90mm,bb=87 -130 437 220]{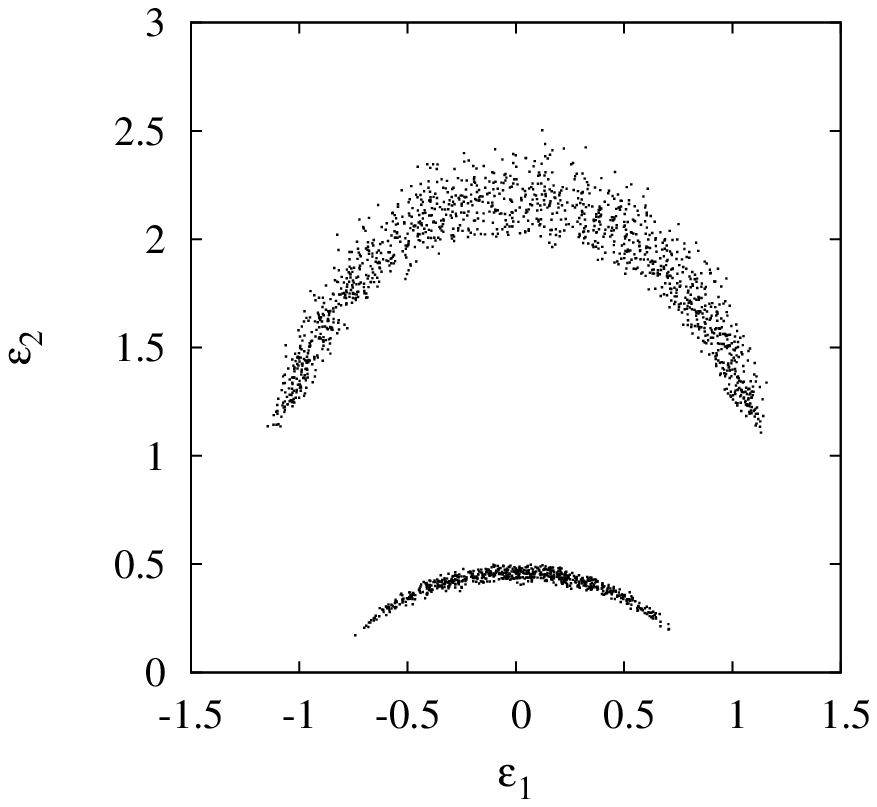}\vspace{-4.7cm}
\caption{\label{fig:fig1} The allowed parameter regions in the
$r_1-r_2$ plane (upper plot) and the $\epsilon_1-\epsilon_2$ plane
(lower plot).}
\end{figure}

We proceed to the numerical illustrations. The input values for the
neutrino parameters are taken from Ref.~\cite{Schwetz:2011qt}. For
example, in the normal hierarchy case, the mass-squared differences
measured in atmospheric and solar neutrino experiments read
\begin{eqnarray}\label{eq:m21}
\Delta m_{21}^2 & =  & \left(7.12 \sim 8.13\right) \times  10^{-5}~ {\rm eV^2}\, , \\
\Delta m_{31}^2 & = & \left(2.18 \sim 2.73\right) \times 10^{-3}~
{\rm eV^2} \, ,
\end{eqnarray}
while the allowed ranges of three mixing angles are
\begin{eqnarray}
\sin^2\theta_{12} & = & 0.27 \sim 0.37 \, , \\
\sin^2\theta_{23} & = & 0.39\sim0.64 \, ,\\
\sin^2\theta_{13} & < & 0.04 \, , \label{eq:t23}
\end{eqnarray}
at $3\sigma$ confidence level. Note that, there are slightly
differences between the fitted parameters in the inverted and normal
hierarchies. In the normal hierarchy case, the above mass-squared
differences correspond to the allowed range of the mass ratio
$5.4<m_3/m_2<5.8$.

In our numerical analysis, we do not make any assumptions on the
model parameters, and randomly choose the values of $r_i$,
$\epsilon_i$ and $m_0$. The predicted neutrino mixing angles and
masses (in the normal hierarchy case) are then compared with
Eqs.~\eqref{eq:m21}-\eqref{eq:t23}, while the allowed parameter
spaces of $r_i$ and $\epsilon_i$ are shown in Fig.~\ref{fig:fig1}.
From the upper plot, one observes that the allowed regions of $r_1$
and $r_2$ are symmetric, which is actually resulted from the
$\nu-\tau$ symmetry in the neutrino mass matrix. In addition, none
of $r_1$ or $r_2$ can be zero, while $r_1 \simeq r_2 \simeq 2$ is
quite favored according to the numerical results. In the lower plot,
$\epsilon_1 =0$ is allowed but $\epsilon_2=0$ is not, indicating
that $\chi$ is required in order to fit the experimental data.
Furthermore, for a fixed value of $\epsilon_1$, there are two
allowed regions for $\epsilon_2$ corresponding to $\epsilon_2
>1$ and $\epsilon_2<1$, respectively.

\begin{figure}[t]
\includegraphics[width=90mm,bb=80 -60 430 290]{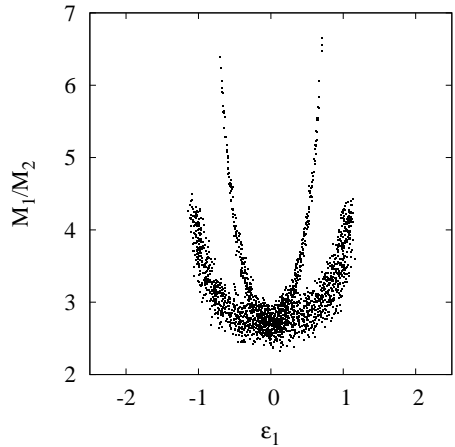}\vspace{-3.cm}
\caption{\label{fig:fig2} The allowed regions of the ratio $M_1/M_2$
with respect to $\epsilon_1$.}\vspace{-.3cm}
\end{figure}

Since the right-handed neutrino masses are also correlated to $r_i$,
we present in Fig.~\ref{fig:fig2} the predicted mass ratio between
two right-handed neutrinos.
One reads from the figure that the mass ratio is generally larger
than 2 showing that the resonant leptogensis
mechanism\cite{Pilaftsis:2005rv} may not simply apply to this
model.~\footnote{The right-handed neutrinos are degenerate in
Ref.~\cite{Hagedorn:2006ug} since their masses are originated from a
bare Majorana mass term, whereas in our model, due to contributions
from $\chi$, a mass splitting between $M_1$ and $M_2$ is included.}

\begin{figure}[t]
\includegraphics[width=90mm,bb=80 -60 430 290]{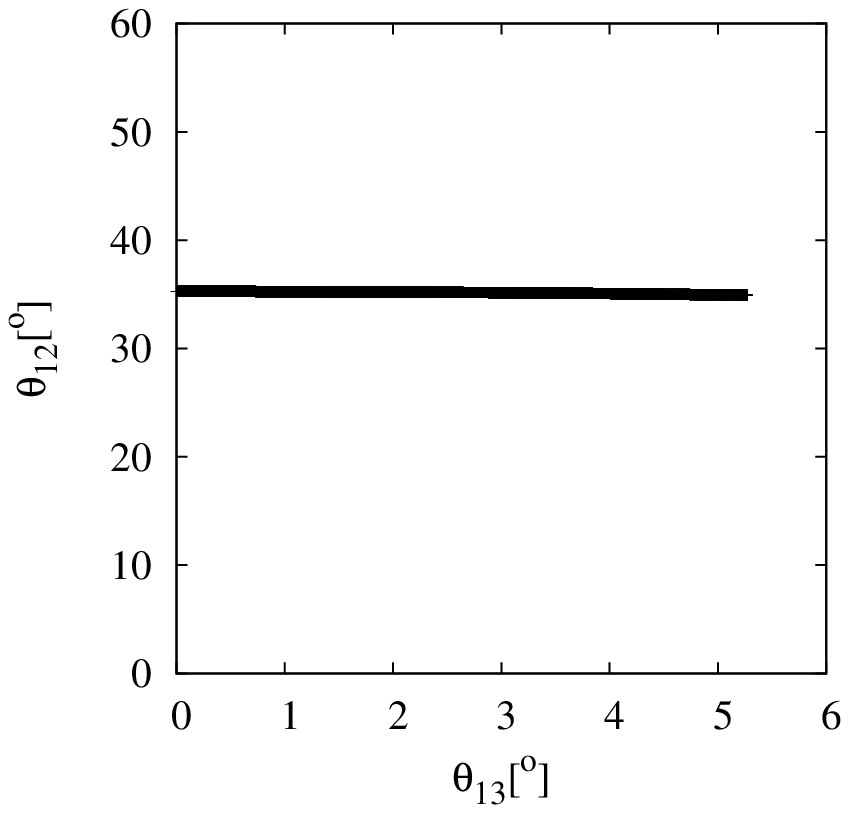}
\includegraphics[width=90mm,bb=80 -160 430 190]{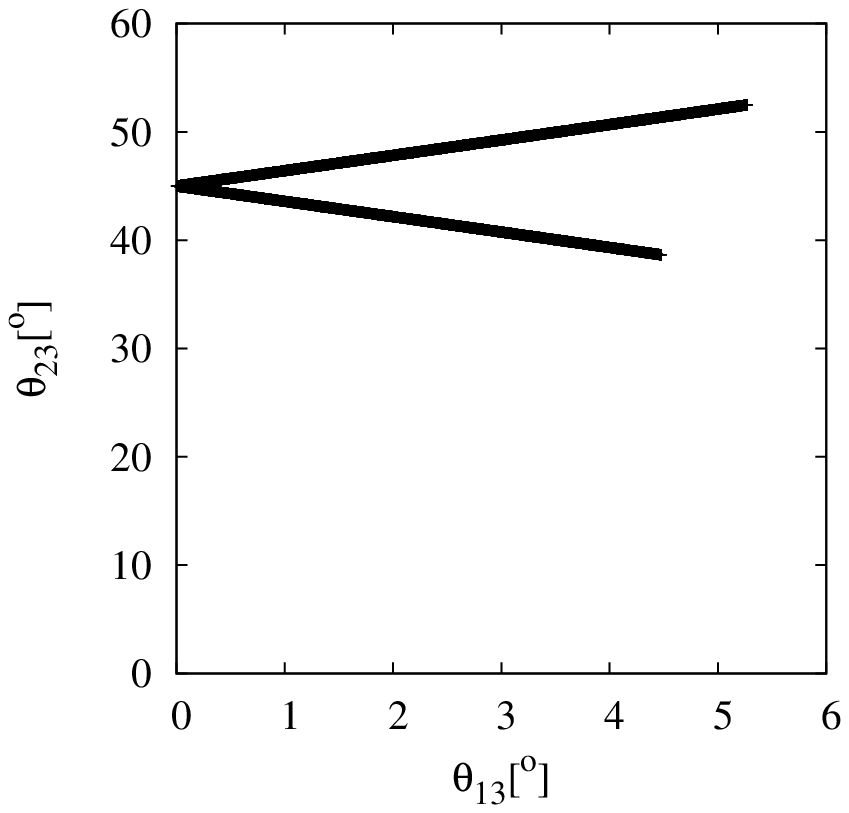}
\includegraphics[width=90mm,bb=85 -260 435 90]{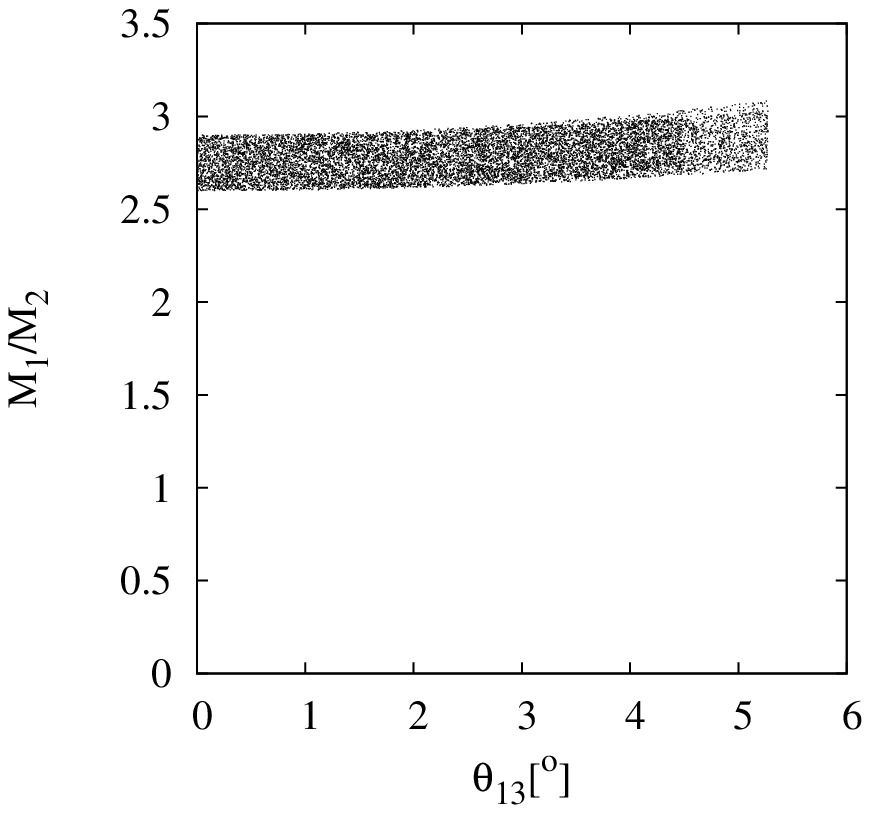}\vspace{-8.2cm}
\caption{\label{fig:fig3} The allowed parameter regions of
$\theta_{ij}$ and $M_1/M_2$.} \vspace{-0.3cm}
\end{figure}

Now we turn to the special case with the assumption $r_1=r_2=2$,
namely, a $\mu -\tau$ symmetry exists in the neutrino sector. The
allowed parameter regions of $\theta_{ij}$ and $M_1/M_2$ are
illustrated in Fig.~\ref{fig:fig3}.
As we expected, there exist strong correlations between three mixing
angles according to the upper and middle plots. This is in good
agreement with our analytical results aforementioned since the
mixing angles are connected by a single parameter $\theta$. The most
severe constraint comes from $\theta_{23}$, and its experimental
allowed range can be fulfilled. As for $\theta_{13}$, an upper bound
$\theta_{13} \lesssim 5^\circ$ can be obtained. As has been shown,
$\theta_{12}$ is confined to it's tri-bimaximal mixing value, and
rather stable compared to the two other mixing angles. In the
particularly interesting limit $\epsilon_1 =0$, the exact
tri-bimaximal mixing pattern will be reproduced. Finally, from the
lower plot, we also find that the right-handed neutrino mass
spectrum should be hierarchical, e.g., $M_1 /M_2 \sim 3$.

\begin{figure}[t]
\includegraphics[width=90mm,bb=80 -60 430 290]{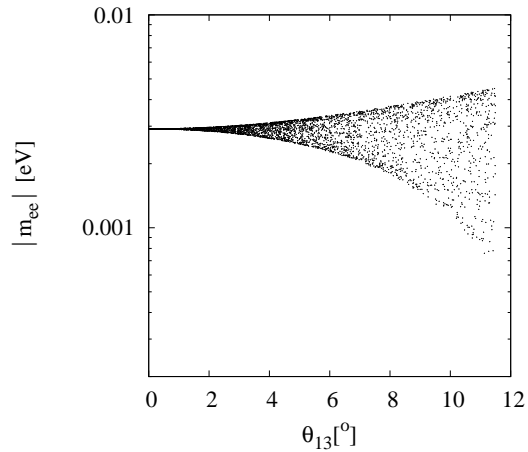}\vspace{-3.cm}
\caption{\label{fig:fig4} The allowed parameter regions of
$|m_{ee}|$ and $\theta_{13}$.} \vspace{-0.cm}
\end{figure}

We stress that our discussions are based on the assumption of real
Yukawa couplings as well as scalar VEVs, whereas in the most general
situation, both of them could be complex. In the presence of
CP-violating effects, the imaginary parts of the model parameters
$A,B,C$ and $X$ could significantly change the predictions addressed
here, and we therefore study it further. Since the neutrinoless
double beta decay process rely on the Majorana feature of light
neutrinos, we illustrate in Fig.~\ref{fig:fig4} the allowed ranges
of the effective mass, i.e., $m_{ee}=\sum m_i U^2_{ei}$, and
$\theta_{13}$ in the presence of CP violation.
The model parameters are the same as those in Fig.~\ref{fig:fig3},
except that we allow them to be complex. One observes from the plot
that any value of $\theta_{13}$ satisfying the current experimental
constraint can be achieved, whereas there exist strong constraints
on $|m_{ee}|$, in particular for a smaller $\theta_{13}$, indicating
potentially attractive signatures in future non-oscillation
experiments.

\section{conclusion}
\label{sec:sec-IV}

In this work, we presented a minimal seesaw model based on the
discrete $S_4$ flavor symmetry. In our model, besides the SM fermion
content, two right-handed neutrinos are introduced transforming as
an $S_4$ doublet. The structure of the model is minimal in the sense
that there are at most two massive light neutrinos which are indeed
required to account for the observed solar and atmospheric neutrino
oscillations. The number of model parameters are reduced greatly
compared to the ordinary type-I seesaw, and thus allow us to make
useful predictions on the neutrino parameters. After carefully
exploring the parameter spaces, we found that the inverted neutrino
mass hierarchy is ruled out whereas the normal hierarchy can be well
accommodated in this framework. In particular, the tri-bimaximal
mixing pattern can be naturally obtained from simple assumptions on
the model parameters, while the deviation of three mixing angles
from their exact tri-bimaximal mixing values are correlated by a
single model parameter. In addition, the right-handed neutrinos
feature a hierarchical mass spectrum, i.e., the ratio between
right-handed neutrino masses is generally larger than 2.

Note that, in the current discussions, we have ignored the
CP-violating effects, since there is yet no direct experimental
information on leptonic CP violation. However, in the most general
case, the CP-violating phases can be easily included since all the
coefficients of Yukawa couplings as well as the VEVs could in
principle be complex. In fact, the CP-violating effects are very
crucial in order to explain the baryon asymmetry of the Universe via
thermal leptogenesis mechanism in the seesaw
models~\cite{Fukugita:1986hr}. In addition, a dirac CP-violating
phase may also be searched for at future long-baseline neutrino
oscillation experiments.

Finally, we stress that the right-handed neutrinos may not be
necessarily heavy, e.g., their masses could be located around keV
scales. One may wonder that, in the mass range $M_i \sim {\rm keV}$
(i.e., the right-handed neutrinos are sterilized), if the
right-handed neutrinos could be viewed as warm dark matter so as to
explain simultaneously the neutrino mass generation and the dark
matter puzzle. Unfortunately, this is not possible in the current
model, since the stability of keV right-handed neutrinos on the
cosmic time scale requires the mixing between sterile and active
neutrinos to be smaller than $10^{-4}$, which leads the mass scale
of light neutrinos to be about $10^{-5}~{\rm
eV}$~\cite{Asaka:2005an}. Such tiny neutrino masses are in conflict
with neutrino oscillation experiments. Possible variations extending
the MSM may provide successful warm dark matter candidate, (e.g., an
additional light right-handed neutrino transforming as a singlet
under $S_4$), which are however beyond the scope of current work.

\begin{acknowledgments}
This work was supported by the ERC under the Starting Grant MANITOP
and by the Deutsche Forschungsgemeinschaft in the Transregio 27
``Neutrinos and beyond -- weakly interacting particles in physics,
astrophysics and cosmology''.
\end{acknowledgments}

\end{document}